\documentclass[pra,superscriptaddress,longbibliography,reprint,floatfix]{revtex4-1}

\usepackage{amsmath}
\usepackage{braket}
\usepackage{graphicx}
\usepackage{epstopdf}
\usepackage{siunitx}

\newcommand{\Hamil}{\hat{\mathcal{H}}}
\newcommand{\Sp}{\hat{S}}

\newcommand{\muB}{\mu_B}

\begin{document}

\title{Counterpart of the Darrieus-Landau instability at a magnetic
  deflagration front}

\author{O. Jukimenko}
\email{oleksii.iukhymenko@umu.se}

\affiliation{Department of Physics, Ume{\aa} University, SE-901\,87 Ume{\aa},
  Sweden}

\author{M. Modestov}

\affiliation{Nordita, KTH Royal Institute of Technology and Stockholm
  University,
Roslagstullsbacken 23, SE-106\,91 Stockholm, Sweden}

\author{C. M. Dion}
\email{claude.dion@umu.se}

\affiliation{Department of Physics, Ume{\aa} University, SE-901\,87 Ume{\aa},
  Sweden}

\author{M. Marklund}
\affiliation{Department of Applied Physics, Chalmers University of Technology,
 SE-412\,96 G\"{o}teborg, Sweden}

\author{V. Bychkov}
\thanks{Deceased}

\affiliation{Department of Physics, Ume{\aa} University, SE-901\,87 Ume{\aa},
  Sweden}

\begin{abstract}
  The magnetic instability at the front of the spin avalanche in a
  crystal of molecular magnets is considered.  This phenomenon reveals
  similar features with the Darrieus-Landau instability, inherent to
  classical combustion flame fronts.  The instability growth rate and
  the cut-off wavelength are investigated with respect to the strength
  of the external magnetic field, both analytically in the limit of an
  infinitely thin front and numerically for finite-width fronts.  The
  presence of quantum tunneling resonances is shown to increase the
  growth rate significantly, which may lead to a possible transition
  from deflagration to detonation regimes.  Different orientations of
  the crystal easy axis are shown to exhibit opposite stability
  properties.  In addition, we suggest experimental conditions that
  could evidence the instability and its influence on the magnetic
  deflagration velocity.
\end{abstract}

\maketitle

\section{Introduction}

The Darrieus-Landau instability, first described in the context of
combustion, is a hydrodynamic instability that is caused by the
thermal expansion of the burning gas~\cite{LL-Fluidmechanics,
  Bychkov_PR_2000, Matalon_ARFM_2007}.  It is characterized by the
fact that the growth rate of the instability at the flame front is
positive for perturbations of any wavelength, and is responsible for
the curving of initially planar flames.  In addition to combustion,
the Darrieus-Landau instability has been observed in different types
of plasmas, from the interstellar medium to inertial confinement
fusion, see, e.g.,
Refs.~\cite{Clavin_PoP_2004,Sanz_PoP_2006,Gamezo_ApJ_2005,Inoue_ApJ_2006,%
  Bychkov_PPCF_2007,Bychkov_PoP_2008,Bychkov_PRL_2011}.

Another system in which combustion-like processes have been observed
are crystals of molecular (nano) magnets.  These molecular magnets
have large spin ($S \sim 10$), and their crystals present an
anisotropy, with an ``easy'' axis along which the spin will align.  In
the presence of an external magnetic field along the easy axis, the
two different orientations will not have the same energy, resulting in
an effective skewed double-well potential (see
\cite{Gatteschi_ACIE_2003} and references therein).  A crystal
prepared in the metastable magnetic orientation, after local heating
to overcome the activation energy, will see a propagation of the spin
reversal, as the energy released by the spin flip will propagate to
neighboring molecules, in a process dubbed a spin avalanche or
magnetic
deflagration~\cite{Suzuki_PRL_2005,Hernandez-Minguez_PRL_2005,%
  Garanin_PRB_2007,Villuendas_EPL_2008,Decelle_PRL_2009}. The spin
reversal can also occur without the activation energy being attained,
through spin
tunneling~\cite{Chudnovsky_Tejada_1998,Delbarco_JLTP_2005}.  This
phenomenon leads to the presence, for certain values of the magnetic
field strength, of tunneling resonances that greatly increase the
speed of propagation of the spin reversal
front~\cite{Hernandez-Minguez_PRL_2005,Garanin_PRB_2007,McHugh_PRB_2007,%
  Macia_PRB_2009,Velez_PRB_2014}.

In previous work~\cite{Jukimenko_PRL_2014}, we demonstrated that the
propagation of this magnetic deflagration front is unstable, due to the
fact that any distortion in the front increases the local magnetic
field, creating a positive feedback.  In this paper, we take a closer
look at the stability of the front, and derive an analytical
expression for the instability growth rate, in the limit of an
infinitely thin front.  We also study the instability numerically,
accounting for a finite magnetic front thickness.  Our results are
also compared to experimental data~\cite{Hernandez-Minguez_PRL_2005},
taking into account the presence of tunneling resonances.

\section{\label{sec:deflagration}Deflagration in crystals of molecular
  nanomagnets}

\begin{figure}
  \centerline{\includegraphics[width=3.3in]{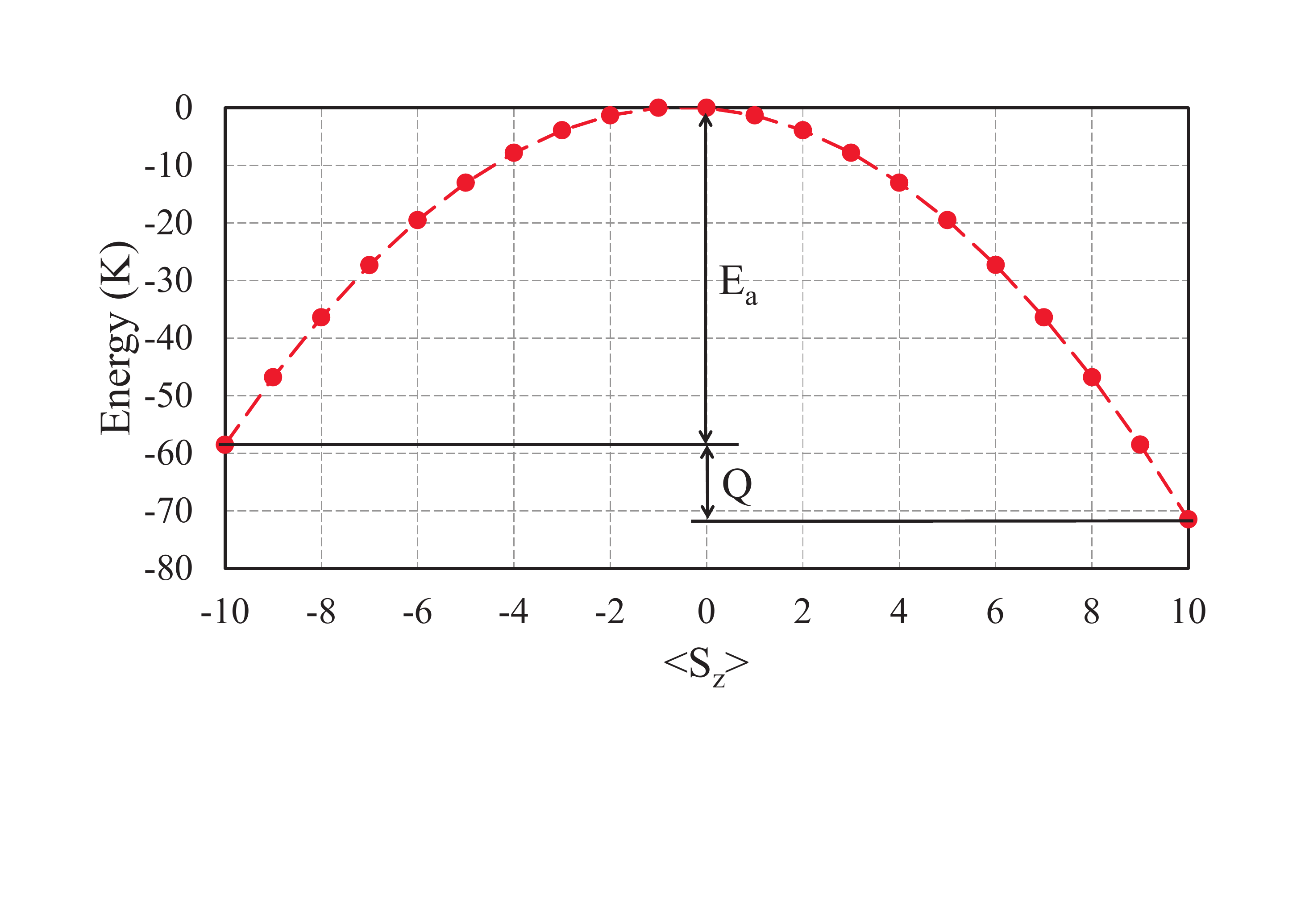}}
  \caption{\label{fig:states}(Color online) Energy levels of a
    molecular magnet Mn$_{12}$-acetate in an external magnetic field
    $B_z = \SI{0.5}{\tesla}$. A molecule initially in the metastable
    state $S_z=-10$ (on the left) must overcome an energy threshold
    $E_a$ in order to relax to the stable state $S_z=10$ (on the
    right). After relaxation, the energy difference (Zeeman energy)
    $Q$ is released as heat.}
\end{figure}

We consider a crystal of Mn$_{12}$-acetate, which has an effective
spin number $S=10$~\cite{Caneschi_JACS_1991}, placed in an external
magnetic field $B_z$ aligned along the $z$-axis, which corresponds
also to the easy axis.  The energy levels of molecular magnet can be
described by the simplified spin Hamiltonian~\cite{Garanin_PRB_2007}
\begin{equation}
  \Hamil = -D \Sp_z^2  - g \muB  B_z \Sp_z,
  \label{eq:hamilton}
\end{equation}
where $D=0.65\ \mathrm{K}$ \cite{Delbarco_JLTP_2005}, $g=1.93$ is the
gyromagnetic factor~\cite{Sessoli_JACS_1993}, and $\muB$ is the Bohr
magneton.  The first term is due to the anisotropy of the crystal,
while the second term describes the dipole interaction between the
external magnetic field and the spin of the molecule.  We consider a
crystal with all molecules initially in the $S_z=-10$ metastable
state, which is then locally heated at one extremity, and study the
propagation of the spin reversal to the stable $S_z=10$ state.  Using
Hamiltonian~(\ref{eq:hamilton}), we find the Zeeman energy release $Q$
\begin{equation}
  Q=2 g \muB B_z S,
  \label{eq:Q}
\end{equation}
and the energy barrier or activation energy $E_a$,
\begin{equation}
  E_a=D S^2 -g \muB  B_z S + \frac{g^2}{4D}\muB^2 B_z^2,
  \label{eq:Ea}
\end{equation}
expressed in temperature units per molecule.  For the particular
external field $B_z = \SI{0.5}{\tesla}$, these quantities are depicted
in Fig.~\ref{fig:states} together with the energy levels of
Mn$_{12}$-acetate.

The evolution of the system is governed by the heat transfer and the
dynamics of the molecules in the metastable state.  The Zeeman energy
release is transformed into phonon thermal energy and is described as
\begin{equation}
\label{eq:heat}
\frac{\partial E}{\partial t} = \nabla \cdot (\kappa \nabla E) -  Q
\frac{\partial n}{\partial t},
\end{equation}
where $E$ is the phonon energy and $\kappa$ is the thermal diffusion
constant, which depends on temperature as $\kappa=\kappa_0
T^{-\beta}$.  The number of molecules in the metastable state evolves
according to
\begin{equation}
  \frac{\partial n}{\partial t} = - \Gamma\left(n-n_{\mathrm{eq}}\right) ,
  \label{eq:dndt}
\end{equation}
where $n_{\mathrm{eq}}=\left[1+\exp\left(Q/T\right)\right]^{-1}$ is
the thermal equilibrium concentration~\cite{Modestov_PRB_2011}.  The
prefactor in Eq.~(\ref{eq:dndt}) stands for the thermal relaxation
rate over the potential barrier $E_a$, shown Fig.~\ref{fig:states}.
In the simplest form, it may be written as the Arrhenius law
\begin{equation}
  \Gamma=\Gamma_R\exp \left( -E_a/T\right).
\end{equation}
Generally speaking, the $\Gamma_R$ factor is not a constant, but
depends on both longitudinal and perpendicular components of the
magnetic field.  In addition, the presence of quantum tunneling
resonances can increase the $\Gamma_R$ factor by several orders of
magnitude for certain values of the magnetic
field~\cite{Garanin_PRB_2012}.

In our analysis, as for experimental measurements, it is more
convenient to work with the temperature variable $T$ rather than the
phonon energy $E$.  The molecular magnets must be kept at cryogenic
temperatures in order to observe the spin reversal phenomenon.  The
typical locking temperature for Mn$_{12}$-acetate lies in a region of
a few degrees above absolute zero.  Under such conditions, the phonon
energy is a strong function of temperature~\cite{kittel63,Garanin_PRB_2008},
\begin{equation}
  \label{eq:phonon}
  E=\frac{A\Theta_D}{\alpha+1}\left(\frac{T}{\Theta_D}\right)^{\alpha+1}
\end{equation}
where $A=13\pi^4/5$ is a constant for this particular crystal type,
$\Theta_{D}=\SI{38}{\kelvin}$ is the Debye temperature, and $\alpha=3$
is the dimensionality of space.

\begin{figure}
  \centerline{\includegraphics[width=3.3in]{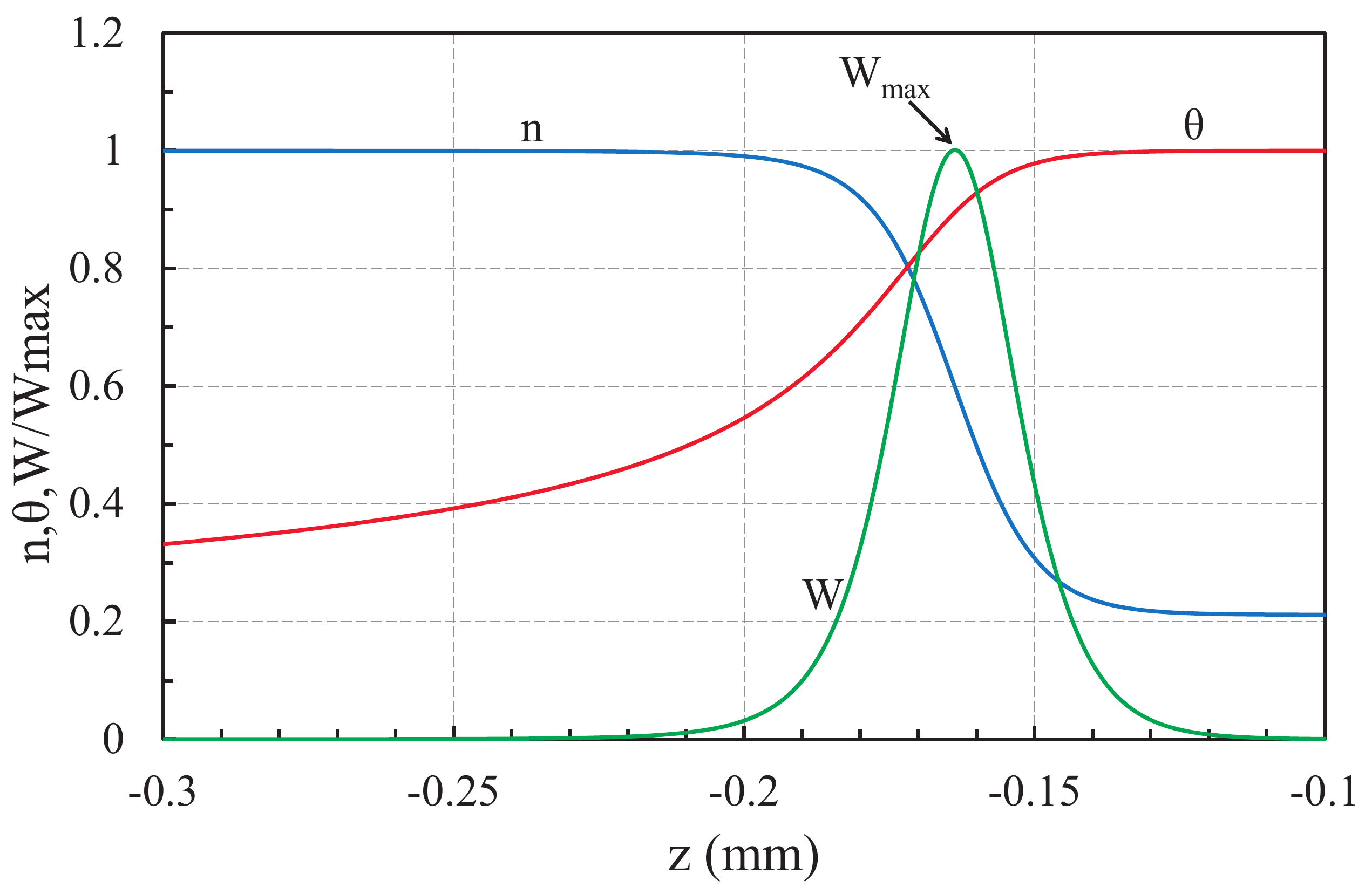}}
  \caption{\label{fig:profile}(Color online) Profile of the stationary
    deflagration front in a crystal of nanomagnets of
    Mn$_{12}$-acetate. The front moves from right to left. The
    external magnetic field is $B_z=\SI{0.5}{\tesla}$ and the final
    temperature is $T_f=\SI{9.9}{\kelvin}$, with $\beta = -13/3$.}
\end{figure}

We start by considering a stationary one-dimensional magnetic
deflagration front, which propagates in the negative $z$ direction
with a velocity $U_f$.  The internal front structure, consisting of
the temperature, energy release, and molecular concentration, is shown
in Fig.~\ref{fig:profile}.
The final temperature $T_f$ behind the front is found from the energy
conservation,
\begin{equation}
  \label{eq:T_f}
  E_0+Q n_0=E_f+Q n_f,
\end{equation}
where index $0$ corresponds to initially ``cold'' matter (left side in
Fig.~\ref{fig:profile}) and index $f$ corresponds to the final ``hot''
(right side) of the front.  In the case of incomplete burning (i.e.,
$n_f\neq0$), this equation is a transcendental one and must be solved
numerically.  We assume that the temperature behind the front $T_f$ is
constant because the heat escape to an external heat sink for this
particular configuration can be neglected. The time for the spin
reversal of the entire sample at the slowest deflagration rate
approximately is $t_d \approx L/U_f =\SI{0.016}{\second} $. The
characteristic time of cooling is $t_c \approx S / \kappa=
\SI{0.19}{\second}$ (for this assumption we have used the thermal
diffusion constant from Ref.~\cite{Velez_PRB_2014},
$\kappa=\SI{0.19}{\meter^2\per\second}$, and surface area
$S=\SI{2.88e-6}{\meter^2}$). Note also that in Ref.~\cite{Velez_PRB_2014},
the time for the sample the return to the temperature of the bath was
measured as $\sim \SI{1}{s}$.  Therefore, since $t_d \ll t_c$, we will
neglect cooling effects in this study.

To simplify further derivations we introduce dimensionless variables
for the coordinate and temperature together with scaled activation and
Zeeman energies,
\begin{equation}
  \xi \equiv zL_f,\quad \theta \equiv T/T_f,\quad \Theta \equiv
  E_a/T_f,\quad \Delta \equiv Q/T_f
  \label{dimless}
\end{equation}
and define
\begin{equation}
  J \equiv \frac{Q\Theta^\alpha_D}{AT^{\alpha+1}_f},\quad
  \kappa \equiv \kappa_0 \theta^{-\beta},\quad
  L_f\equiv\frac{\kappa_0 T_f^{-\beta}}{U_f}.
  \label{denotes}
\end{equation}
Here, $L_f$ is a characteristic length of the problem.  Then in the
reference frame of the moving front the
equations~(\ref{eq:heat})--(\ref{eq:dndt}) form the dimensionless system
\begin{equation}
  \label{eq:Sys_stat}
  \begin{split}
    \frac{d \psi}{d\xi }& =\theta^\beta\psi-J\Lambda b_z n
    e^{-\Theta/\theta}\left(n-n_{\mathrm{eq}}\right),\\
    \frac{d \theta}{d\xi }& =\psi\theta^{\beta-\alpha},\\
    \frac{dn}{d\xi }& =-\Lambda
    e^{-\Theta/\theta}\left(n-n_{\mathrm{eq}}\right),
  \end{split}
\end{equation}
where $\psi$ stands for the heat flux and $\Lambda=L_f\Gamma_R/U_f$ is
an eigenvalue of the stationary front.  The new variable $\psi$ allows
us to write down the governing system as a set of first-order
differential equations, which is important for stability analysis
described in Sec.~\ref{sec:stability}.  In order to calculate the
stationary profiles depicted in Fig.~\ref{fig:profile}, we integrate
the system~(\ref{eq:Sys_stat}) from the left, ``cold'' side towards
the right, ``hot'' side, and the eigenvalue $\Lambda$ is found by this
shooting method~\cite{Modestov_PRE_2009}, matching the results of
numerical integration to the analytical solution given by
Eq.~(\ref{eq:T_f}) for the final temperature.

It should be noted that the assumption of a stationary front is valid
when the front thickness is much smaller than the length of the
sample.  The characteristic front width can be determined as the
half-width of the energy release peak; from Fig.~\ref{fig:profile} we
find that $L_{f}\approx \SI{0.025}{\milli\meter}$.  The typical
crystal size used in experiments is
1--\SI{2}{\milli\meter}~\cite{Velez_PRB_2014}, which is almost two
orders of magnitude larger than the front width.  Consequently, there
is enough room to form a steady propagating front of magnetic
deflagration.

\begin{figure}
  \centerline{\includegraphics[width=3.3in]{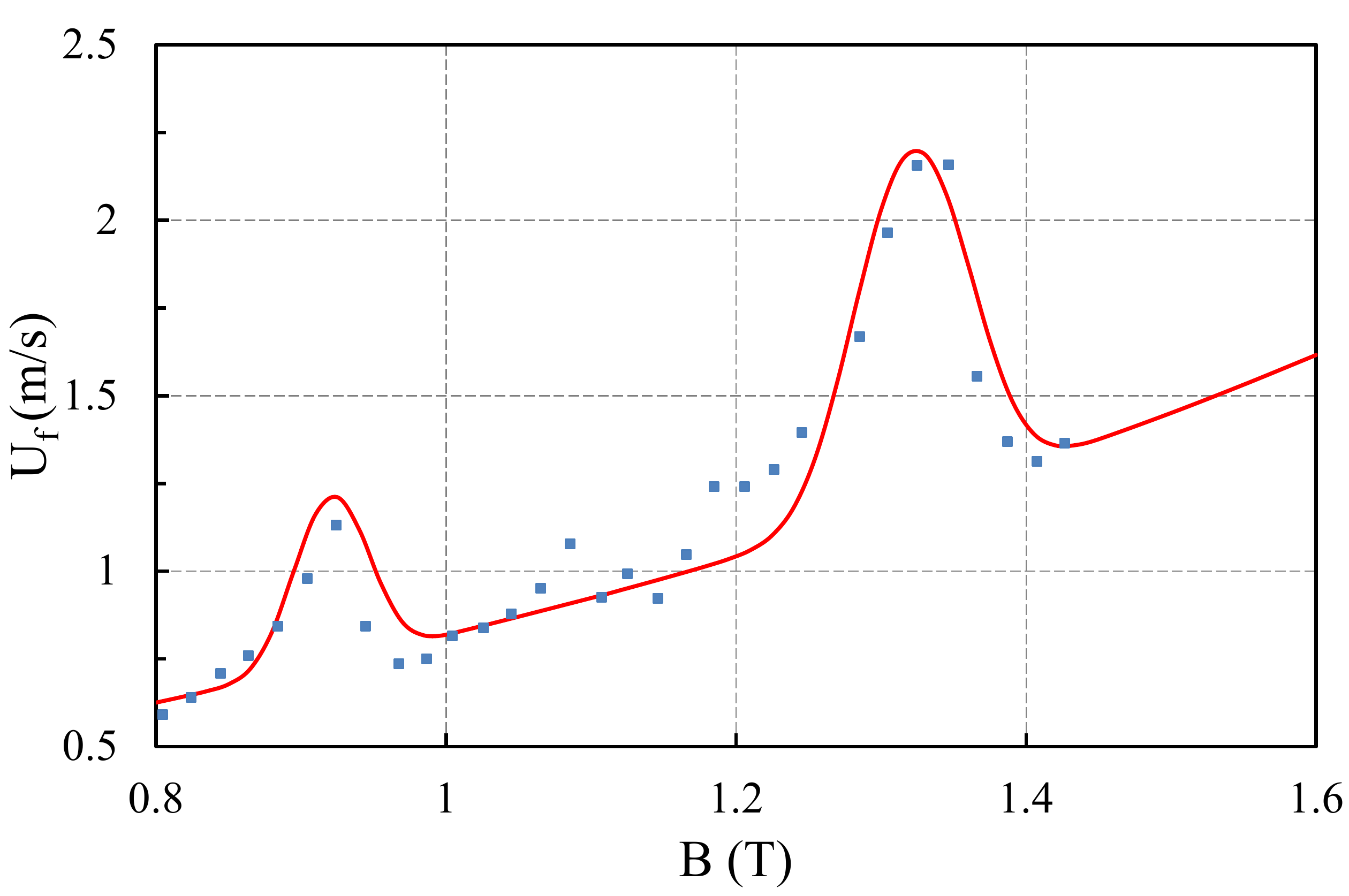}}
  \caption{\label{fig:res_vel}(Color online) Magnetic deflagration
    velocity $U_f$ for the planar front as a function of the
    longitudinal magnetic field $B_z$.  The two peaks occur due to the
    tunneling resonance.  The markers correspond to experimental data
    extracted from Ref.~\cite{Hernandez-Minguez_PRL_2005} and the line
    represents the fitted theoretical dependence.}
\end{figure}

Another issue to mention is the ambiguity in determining the front
velocity $U_f$.  Resolving the stationary profiles, we compute the
front eigenvalue $\Lambda$, however in order to find $U_f$ we need to
know the value of $\kappa_0 \Gamma_R$ according to expressions
(\ref{denotes}), as
\begin{equation}
  \Lambda = \frac{\kappa_0 \Gamma_R T_f^{-\beta}}{U_f^2}.
\end{equation}
So far, the actual parameters $\kappa_0$ and $\Gamma_R$ have not been
measured, and we estimate the relation $\kappa_0 \Gamma_R$ by fitting
the velocity to experimental data~\cite{Hernandez-Minguez_PRL_2005}.
The dependence of $\Gamma_R$ on the magnetic field has been the
subject of many studies, see, e.g.,
Refs.~\cite{Leuenberger_PRB_2000,Garanin_PRB_2010}. In this paper, we
interpolate $\Gamma_R(B)$ by fitting experimental
data~\cite{Hernandez-Minguez_PRL_2005} (see Fig.~\ref{fig:res_vel})
using a Gaussian function to model tunneling resonances
as~\cite{Jukimenko_PRL_2014}
\begin{equation}
  \Gamma_R(B)=\Gamma_0 \left\{1+\sum_i a_{i}\exp\left[-b_{i}\left(\frac{B}{B_{i}}-1\right)^2\right]\right\},
\label{GammaR}
\end{equation}
where $B_i$ is the resonance magnetic field, $\Gamma_0$ is a constant,
and parameters $a_i$, $b_i$ are the amplitude and the width of the
resonance, respectively.  According to experimental
data~\cite{Hernandez-Minguez_PRL_2005} shown in
Fig.~\ref{fig:res_vel}, these parameters are calculated as
\begin{equation}
  \label{ab_res}
  \begin{aligned}
    B_1 &= \SI{0.92}{\tesla}, & a_1&=1.89, & b_1&=840, \\
    B_2 &= \SI{1.32}{\tesla}, & a_2&=2.61, & b_2&=870,
  \end{aligned}
\end{equation}
with the estimate $\kappa_0\Gamma_0\approx
\SI[per-mode=symbol]{4e5}{\meter\squared\per\second\squared}$.

\section{\label{sec:analytical}Analytical instability analysis within infinitely thin front}

\begin{figure}
  \centerline{\includegraphics[width=3.3in]{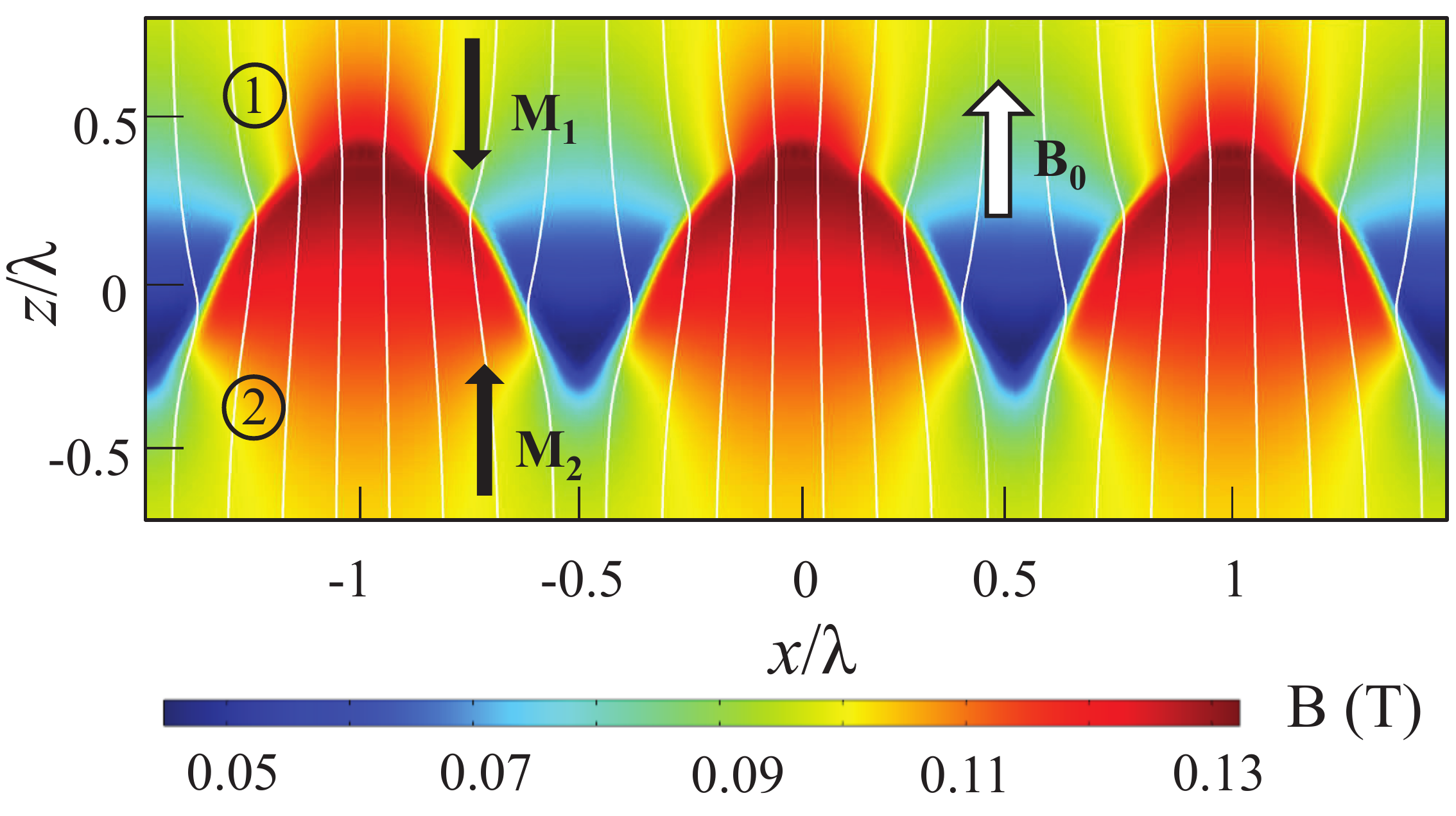}}
  \caption{\label{fig:instability}(Color online) Magnitude of the
    magnetic field in a 2D simulation for magnetic deflagration with a
    corrugated front. The external magnetic field is parallel to the
    $z$-axis, $B_z=\SI{0.5}{\tesla}$. The front moves along the
    $z$-axis in the positive direction. The magnetization of the
    medium flips from $\mathbf{M}_1=(0;-M)$ in region 1 to
    $\mathbf{M}_2=(0;M)$ in region 2, with
    $\mu_0M=\SI{0.05}{\tesla}$. The dipole field produced by the
    crystal results in an increase of the field at the tip of the
    humps.}
\end{figure}

The propagation of the magnetic deflagration front is
unstable~\cite{Jukimenko_PRL_2014} as any distortion of the front
increases the magnetic field where the front bends, as shown in
Fig.~\ref{fig:instability}. This, in turn, leads to an increase of
the front velocity, resulting in positive feedback.  In this section,
we will take a closer look at the front stability properties.

In order to perform this analysis we need to make two assumptions.
First, we assume that the magnetization of the particular nanomagnet
is produced by the spins of the molecules, with the behavior of the
spins described by Hamiltonian~(\ref{eq:hamilton}).  The external
magnetic field will significantly affect the eigenstates of
Hamiltonian only when the second term in Eq.~(\ref{eq:hamilton})
becomes comparable to the first term, i.e., for an anisotropy field
$B_A=2D/(g \mu_B)$.  For Mn$_{12}$-acetate, $B_A$ is of the order of
\SI{10}{\tesla}~\cite{Garanin_PRB_1997}.   In our study, we focus on
fields strengths much lower than the anisotropy field, where the dependence of
$\mathbf{M}$ on $\mathbf{B}$ is negligible, see Fig.~2
in Ref.~\cite{Dion_PRB_2013}. Therefore, we suppose that the amplitude of the
crystal magnetization $\mathbf{M}$ does not depend on the strength of
the external field.  Second, we assume that front width is infinitely
thin, so that the profiles presented in Fig.~\ref{fig:profile} reduce
to step functions, separating the cold and the hot regions of the
crystal.  We will remove these restrictions in the next section and
consider the instability properties for a continuous front structure.

\begin{figure}
  \centerline{\includegraphics[width=3.3in]{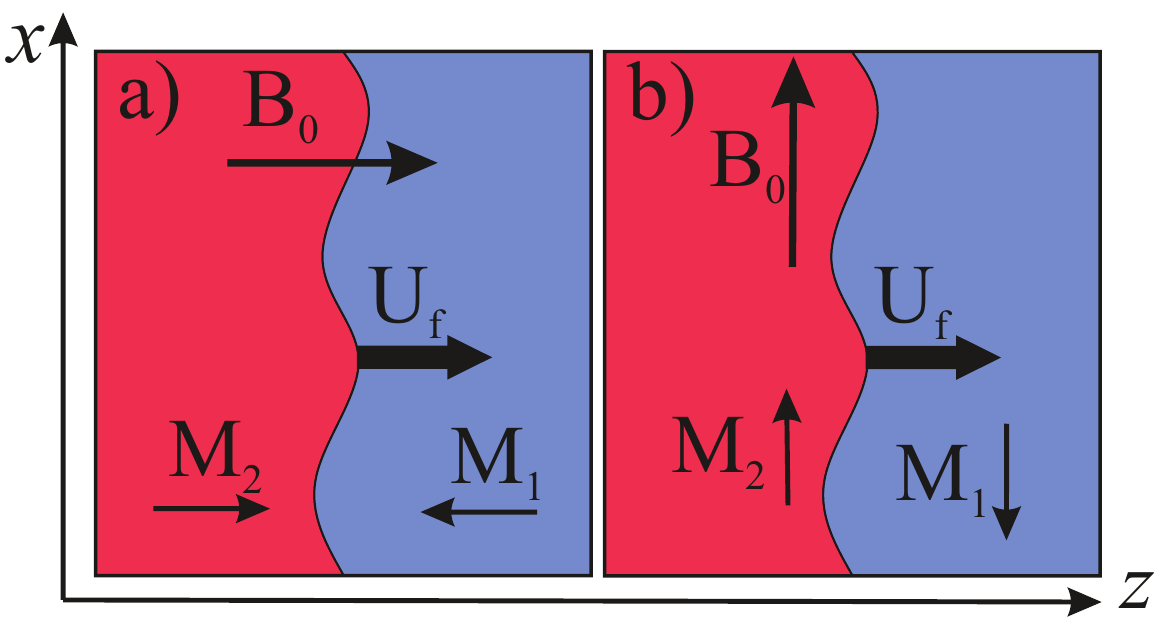}}
  \caption{\label{fig:linear_geometry}(Color online) Possible
    orientations of the front propagation vs.\ the crystal easy axis:
    (a) Most common geometry used in experiments, where the easy axis is
    parallel to the direction of the front propagation; (b) Hypothetical
    orientation of the crystal, where the easy axis is perpendicular to
    the direction of the front propagation.  The external magnetic field
    $B_0$ is oriented along the easy axis of the crystal.}
\end{figure}

With respect to the front propagation and easy axis, multiple mutual
orientations are possible. We consider the two principal cases: the front
propagation is aligned with, Fig.~\ref{fig:linear_geometry}(a), or
perpendicular to, Fig.~\ref{fig:linear_geometry}(b),
the easy axis of the crystal.
The position of a planar front, propagating with
a constant velocity $U_f$, is given by $Z_f=U_ft$.  The front is then
perturbed with a superposition of Fourier modes written as
\begin{equation}
  \label{eq:Zpert}
  Z_f(x,t) =U_ft + \sum_k\widetilde{Z}_k\exp(ikx+\sigma t),
\end{equation}
where $\sigma$ is the instability growth rate, $k=2\pi/\lambda$ is the
wave number, and $\lambda$ is the wavelength of the perturbations.  If
$\mathrm{Re}(\sigma)>0$ the perturbations grow in time and the front
becomes unstable; in the opposite case, $\mathrm{Re}(\sigma)<0$, the
front is remains stable.  The imaginary part of $\sigma$ leads to
oscillations and pulsations of the front~\cite{Modestov_PRB_2011}; in
this paper we consider the case when $\mathrm{Im}(\sigma)=0$, such
that $\sigma$ takes only purely real values.

We start with the case when the front propagates parallel to the easy
axis and the magnetic field, Fig.~\ref{fig:linear_geometry}(a).  This
is a common crystal orientation in experimental
studies~\cite{McHugh_PRB_2007, Velez_PRB_2014, Decelle_PRL_2009}.
Magnetization flips from $\mathbf{M}_1=(0;-M)$ in the cold region
before the front to $\mathbf{M}_2=(0;M)$ in the hot region behind the
front, see Fig.~\ref{fig:linear_geometry}(a).  The deformation of the front
induces perturbations in the magnetic field,
\begin{equation}
  \mathbf{B}_z=\mathbf{B}_0+\sum_k\widetilde{\mathbf{B}}_k(z)\exp(ikx+\sigma t).
\end{equation}
The magnetic field inside the crystal is governed by the stationary
Maxwell equations for nonconducting media,
\begin{align}
  \label{eq:Maxwell}
  \nabla\times\mathbf{H}&=0, & \nabla\cdot\mathbf{B}&=0,
\end{align}
with the relation $\mathbf{B}/\mu_0=\mathbf{H}+\mathbf{M}$, where
$\mu_0$ is the magnetic constant.  Far from the front all the
perturbations must vanish, such that the magnetic field perturbations
along $z$ can be written as $\widetilde{\mathbf{B}}_{1,2}(z) \propto
\exp(\mp kz)$.  Boundary conditions for the magnetic field on the
front interface are
\begin{align}
  \widehat{\mathbf{a}}_{n}\cdot
  \left[\mathbf{B}\right] &= 0, &
  \widehat{\mathbf{a}}_{n}\times \left[\mathbf{H}\right] &= 0,
\end{align}
where $\left[F\right] \equiv F_{2}-F_{1}$ designates the difference of
any value $F$ across the front and the normal vector to the perturbed
front is $\widehat{\mathbf{a}}_{n} \equiv
\widehat{\mathbf{a}}_{z}-\widehat{\mathbf{a}}_{x}\partial_{x}{Z}_{f}$.

Resolving Maxwell's equations~(\ref{eq:Maxwell}) together with the
boundary conditions, we find the relation between the magnetic fields
ahead and behind the front.  Taking $z=0$ we find that
\begin{equation}
\widetilde{B}_{z1}= \widetilde{B}_{z2}= \mu_{0}M k
\widetilde{Z}_{f},
\end{equation}
which leads to an increase of the magnetic field at the tip of the
hump.  Within the linear stability problem, the perturbation of the
front velocity is given by $\partial_{t} \widetilde{Z}_{f} = U'_{f}
\widetilde{B}_{z}$, where $U'_{f}\equiv
dU_{f}/dB$~\cite{Jukimenko_PRL_2014}.  It yields the dispersion
relation in a very concise form as
\begin{equation}
  \label{eq:disp1}
  \sigma =k U'_{f} \mu_{0} M.
\end{equation}
This result means that an infinitely thin magnetic deflagration front
is unstable with respect to perturbations of all wavelength, since
$\sigma>0$ for any $k$.  Mathematically, this relation $\sigma\propto
k$ is similar to the Darrieus-Landau
instability~\cite{Bychkov_PR_2000,Matalon_ARFM_2007}.

Next, we consider the crystal configuration where the front propagation is
perpendicular to the easy axis of the crystal,
Fig.~\ref{fig:linear_geometry}(b).  In this case, the magnetization varies
as $\mathbf{M}_{1,2}=(\mp M;0)$ and the external magnetic field is given
as $\mathbf{B}_0=\hat{\mathbf{a}}_x B_0$.  Using the same approach as
above, we obtain the dispersion relation
\begin{equation}
  \label{eq:disp2}
  \sigma =-k U'_{f} \mu_{0} M.
\end{equation}
Hence such a configuration results in a stable propagation of the
magnetic deflagration wave.

Characteristic values of the relative strength of the instability
$\sigma/U_{f}k$ may vary significantly for different materials,
depending on the magnetization $M$ and front velocity sensitivity
$U'_{f}$.  We therefore expect noticeable magnetic instabilities in
two cases, when either $M$ or $U'_{f}$ are high.  Strong magnetization
can be found in ferromagnetic materials, so this instability might
affect propagation of the domain walls.  In magnetic nanomagnets, the
magnetization is relatively weak, $\mu_{0}M \sim
\SI{0.05}{\tesla}$~\cite{Garanin_PRB_2012}, but the velocity slope
$U'_{f}$ theoretically can reach infinite values at the tunneling
resonances~\cite{Garanin_PRB_2013}.

\section{\label{sec:stability}Stability analysis accounting for the
  internal front structure}

In Sec.~\ref{sec:analytical}, we found that the magnetic deflagration
front is unstable in the infinitely thin front limit.  However, such a
method does not provide any characteristic length scale for the
instability nor the strength of the instability or its dependence on
the external magnetic field.  Here, we investigate the instability
properties taking into account a finite front width and the continuous
structure of the deflagration front obtained in
Sec.~\ref{sec:deflagration}.

For a finite front thickness, the magnetic field $\mathbf{B}$, together
with the all other variables, changes continuously within the front.
We introduce the magnetic vector potential $\mathbf{A}$ defined from
$\mathbf{B}=\nabla\times\mathbf{A}$, such that the first Maxwell
equation $\nabla\cdot\mathbf{B}=0$ is satisfied automatically.  For a
planar front, the vector potential has only one component
$\mathbf{A}=(0,A(x,z),0)$ and for uniform field $B_0$ it reduces to
$\mathbf{A}=(0,xB_0,0)$.  Consequently, the magnetic field components
are
\begin{align}
  \label{eq:magvec1}
  B_x &= -\frac{dA}{dz}, & B_z &= \frac{dA}{dx}.
\end{align}
The second Maxwell equation $\nabla\times\mathbf{H}=0$ can be
rewritten as
\begin{equation}
  \label{eq:magvec2}
  \nabla^2A-2\mu_0M_0\frac{dn}{dx} = 0,
\end{equation}
where we assume that the magnetization changes proportionally to the ratio
of the metastable molecules $\mathbf{M}=(0,0,M_0(2n-1))$.  We
introduce the dimensionless magnetic potential defined as
$a \equiv A/(L_f\mu_0M_0)$ and a new variable 
$\chi\equiv i \, da/d\xi$
in order to have differential equations of the first order only.

Next, we apply a small perturbation so that every variable is written
in as $f(z,x,t)=f(z)+\tilde{f}(z)\exp(ikx+\sigma t)$.  After
straightforward calculations, the linearized equations
(\ref{eq:Sys_stat}) and (\ref{eq:magvec2}) can be written in  matrix
form as
\begin{equation}
  \label{eq:fullsys}
  \frac{d}{d\xi}\mathbf{v}=\mathbf{D}\mathbf{v},
\end{equation}
where $\mathbf{v}=\left(\tilde{\theta},
  \tilde{n},\tilde{\psi},\tilde{a},\tilde{\chi}\right)^{T}$ is the
vector of perturbations. $\mathbf{D}=\mathbf{D}\left(\xi,S,K\right)$
is a $5\times5$ matrix of the coefficients of the system of differential
equations,
\begin{equation}
  \label{eq:D_matrix}
  \mathbf{D}=\begin{pmatrix}
    (\beta-\alpha)\psi\theta^{\beta-\alpha-1} & 0& \theta^{\beta-\alpha} &0&0\\
    -\Phi & -S-W & 0& D_{24} & 0\\
    D_{31} & -WJ & \theta^\beta & D_{34} & 0\\
    0 & 0 & 0 & 0 & 1 \\
    0 & -2K & 0 & K^2 & 0\\
\end{pmatrix},
\end{equation}
where the matrix components are
\begin{equation}
  \begin{aligned}
    \label{eq:D_coeff}
    D_{24} &\equiv K\left[
      W\frac{C_1}{\theta}(n-n_{\mathrm{eq}})-\frac{C_2}{b_{0z}}\right], \\
    D_{31} &\equiv \beta\psi\theta^{\beta-1}-JW\frac{\Theta}{\theta^2
    }\left(n-n_{\mathrm{eq}}-n^2_{\mathrm{eq}}\frac{\Delta}{\Theta}
      e^{\Delta/\theta}\right) \\
    &\quad +S\theta^\alpha+K^2\theta^{\alpha-\beta}, \\
    D_{34} &\equiv \frac{K}{b_z} \left[ W(n-n_{\mathrm{eq}}) \left(
        1+b_{0z}\frac{C_1}{\theta} \right) -C_2 \right],
  \end{aligned}
\end{equation}
with
\begin{align*}
  W &\equiv \Lambda e^{-\Theta/\theta}, &
  C_1 &\equiv \frac{g \mu_B S_z}{T_f \mu_0 M_0}, &
  C_2 &\equiv W n^2_{\mathrm{eq}}
\end{align*}
introduced for brevity.  In the equations above, $b_{0z} \equiv \mu_0
M_0B_{0z}$ is a dimensionless external magnetic field, and $S \equiv
\sigma\Gamma_0$ and $K \equiv kL_f$ are the scaled instability growth rate
and perturbation wave number, respectively.  Some of the matrix
coefficients in~(\ref{eq:D_matrix}) are known from the stationary
profile, while the others depend on $S$ and $K$ as parameters.

In order to find the instability dispersion relation $S(K)$, we apply
the same method as in similar studies of
instabilities~\cite{Modestov_PRE_2009,Modestov_PRB_2011}.  First, we
search for solutions to the system~(\ref{eq:fullsys}) in the uniform
regions, where all the coefficients in matrix $\mathbf{D}$ are
constant.  In this case the perturbations decay exponentially as
\begin{equation}
  \label{eq:prtrdec}
  \lim_{\xi\rightarrow\pm\infty} \mathbf{v} = \mathbf{v}_i
  \exp{\left(\mu_i\xi\right)},
\end{equation}
where the $\mu_i$ are the so-called system modes and $\mathbf{v}_i$
are constant perturbation amplitudes.  Substituting
Eq.~(\ref{eq:prtrdec}) into Eqs.~(\ref{eq:fullsys}), we obtain
\begin{equation}
  \label{eq:eig}
  \mathbf{D}\mathbf{v}=\mu\mathbf{v}.
\end{equation}
We compute the eigenvalues $\mu_i$ and the corresponding eigenvectors
$\mathbf{v}_i$.  We consequently obtain five modes for the cold and
the hot sides of the front, although not all of them are physical.  In
order to pick out the physical eigenvectors, we use the condition that
perturbations must vanish far from the front as $
\lim_{\xi\rightarrow\pm\infty} \mathbf{v} \rightarrow 0$. In other words,
we consider eigenvectors $\mathbf{v}_i$ for which $\mu_i > 0$ at
$\xi\rightarrow-\infty$ or $\mu_i < 0$ at $\xi\rightarrow+\infty$.
If the problem is self-consistent, there will be exactly five modes
$\mu_i$ satisfying these conditions, usually three modes on
one side and two on the other.

After that we integrate Eqs.~(\ref{eq:fullsys}) from the front
boundaries using $\mathbf{v}_i$ as boundary conditions.  We match the
results of the integration at the point of maximal energy release
$W_{\mathrm{max}}$ (shown in Fig.~\ref{fig:profile}).  Generally
speaking this matching point can be chosen in a different way without
affecting the final results, however the current choice minimizes the
numerical integration errors~\cite{Modestov_PRE_2009}.  At this point
the integrated amplitudes constitute a matrix and the dispersion
relation $S=S(K)$ is obtained when the matrix determinant becomes
equal to zero.

\section{Results and discussion}

\begin{figure}
  \centerline{\includegraphics[width=3.3in]{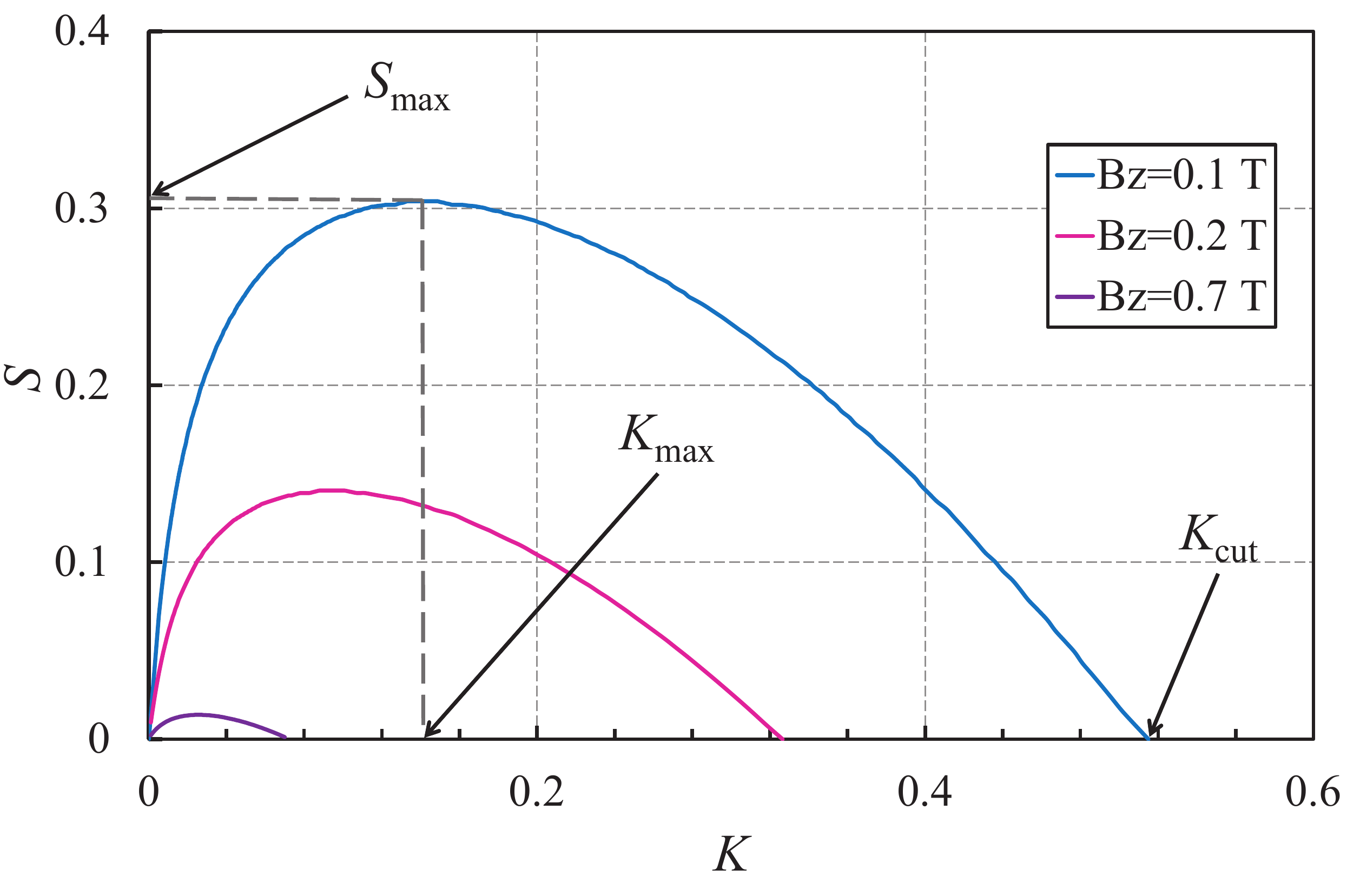}}
  \caption{\label{fig:dispersion}(Color online) Dimensionless
    instability growth rate versus the dimensionless wave number for
    different magnetic fields. Values $K_{\mathrm{cut}}$,
    $K_{\mathrm{max}}$, and $S_{\mathrm{max}}$ are shown only for the
    dispersion relation for $B_z = \SI{0.1}{\tesla}$.}
\end{figure}
The dispersion relation $S=S(K)$ is shown in
Fig.~\ref{fig:dispersion}, for several magnitudes of external magnetic
field.  It has a parabola-like shape similar to the one obtained for
the Darrieus-Landau instability in combustion and laser
ablation~\cite{LL-Fluidmechanics,Bychkov_PR_2000,Modestov_PRE_2009}.
In the region of small wave numbers the instability displays a strong
increase of the growth rate against the variation of the wave number.
Then, at a certain wave number $K_{\mathrm{max}}$, the instability
growth rate reaches its maximum $S_{\mathrm{max}}$.  After that, the
instability becomes weaker until it vanishes at $K_{\mathrm{cut}}$.
As in the case of the Darrieus-Landau instability, the stabilization
is attributed to the final front width due to thermal conduction.

Another important outcome from Fig.~\ref{fig:dispersion} is that the
instability is stronger for relatively weak fields.  This can be
qualitatively explained in the following way.  The instability is
caused by the dipole field created by the crystal magnetization.  In
our model this magnetization does not depend on the external magnetic
field.  Hence for weak fields ($B_0<\SI{0.1}{\tesla}$) the
magnetization at the curved front creates a relatively strong dipole
field as compared to the external field.  This, in turn, increases the
front velocity driving the instability to grow further.  On the other
hand, for high external fields, the increase due to crystal
magnetization is relatively weak, leading to a much smaller increase
of the front velocity.

\begin{figure}
  \centerline{\includegraphics[width=3.3in]{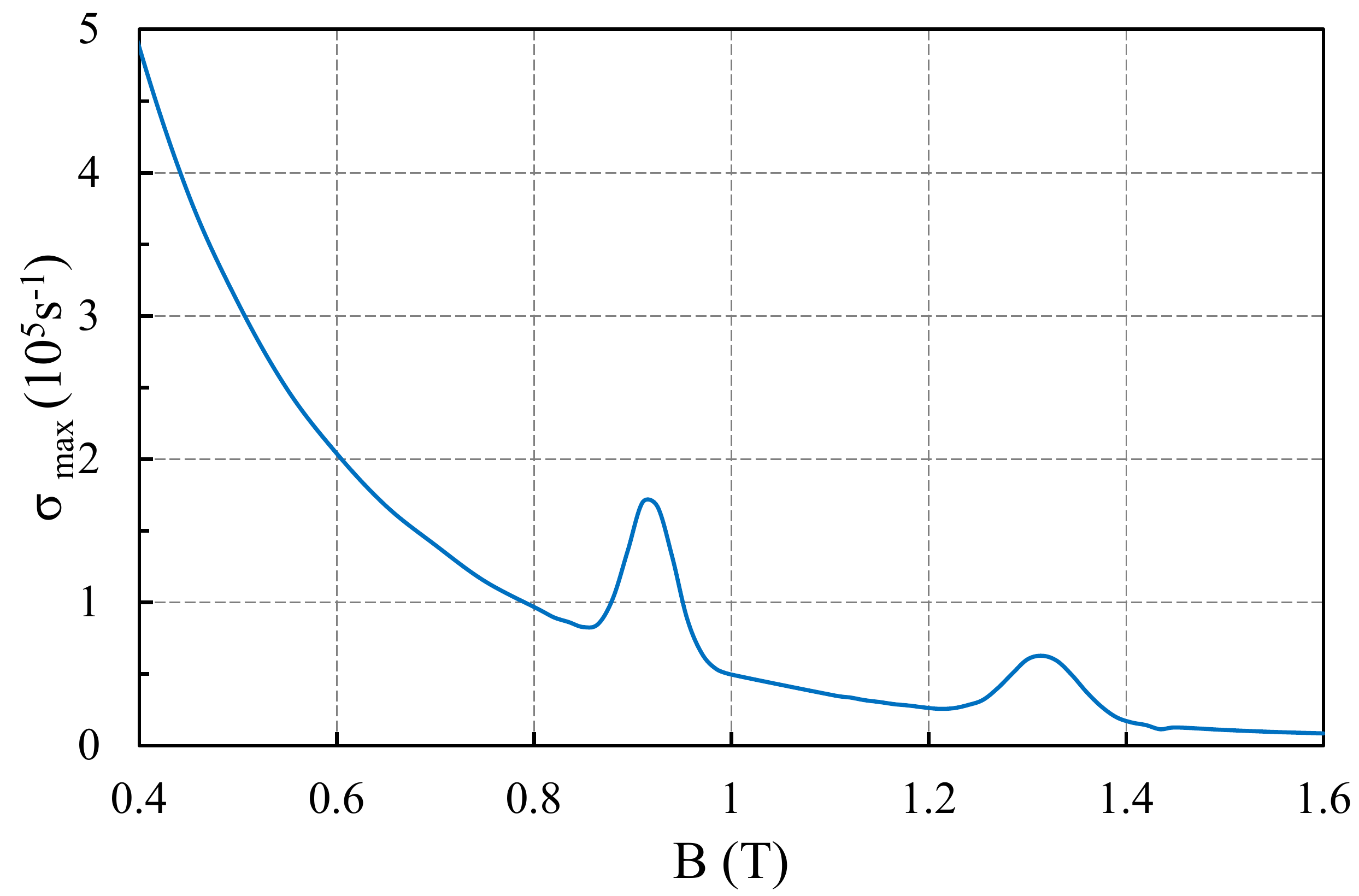}}
  \caption{\label{fig:Smax}(Color online) Maximum of the instability
    growth rate versus magnetic field for Mn$_{12}$-acetate.}
\end{figure}

\begin{figure}
  \centerline{\includegraphics[width=3.3in]{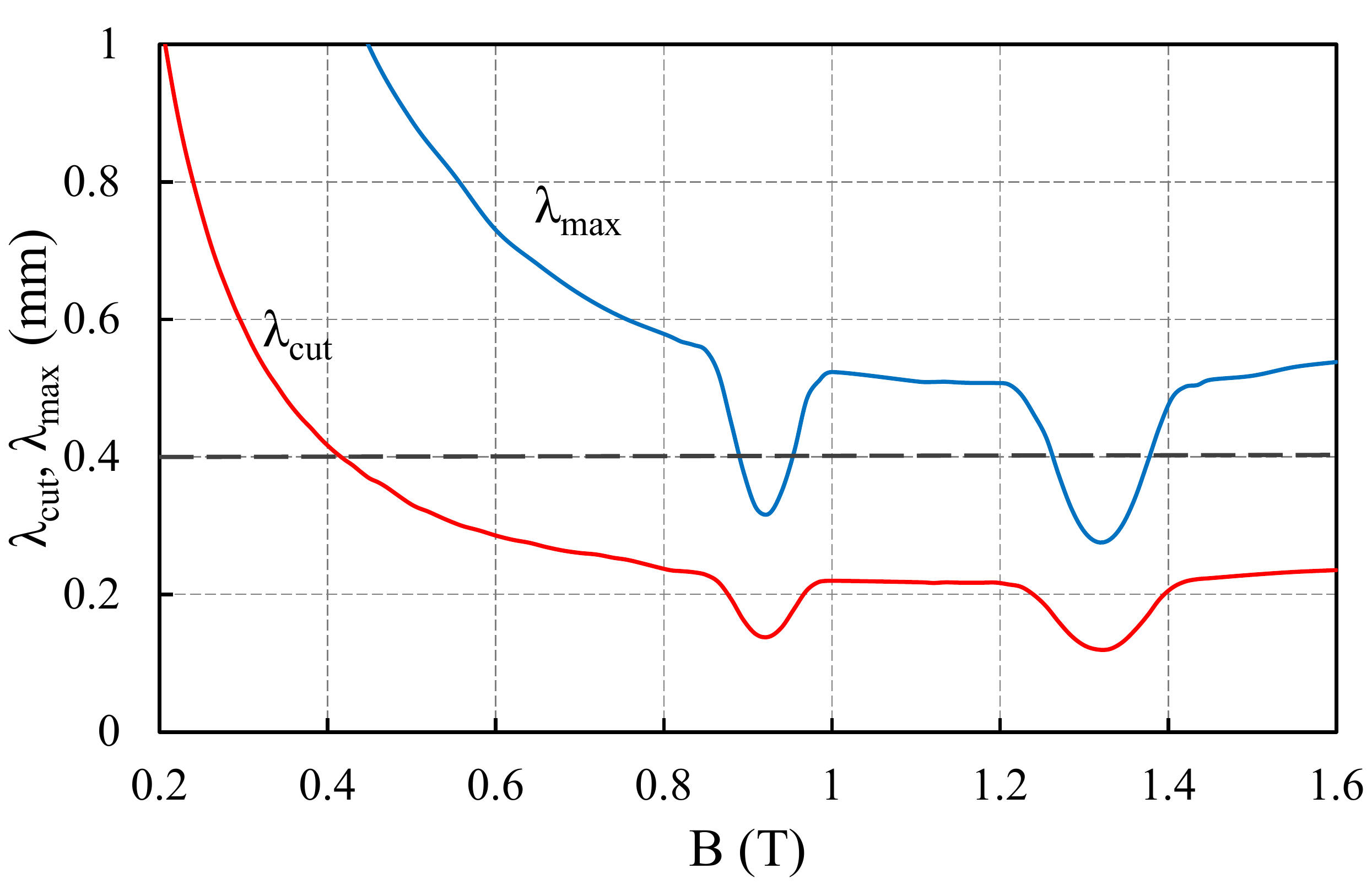}}
  \caption{\label{fig:cutoff}(Color online) Cutoff wavelength and
    wavelength at the maximum, for Mn$_{12}$-acetate. The horizontal
    dashed gray line shows the typical sample width
    \SI{0.4}{\milli\meter} in experiments~\cite{Velez_PRB_2014}.}
\end{figure}

The influence of the magnetic field is better shown in
Figs.~\ref{fig:Smax} and~\ref{fig:cutoff}, where we present the
maximum instability growth rate as well as the cutoff wavelength and
wavelength at the maximum as a function of the external magnetic
field.  In both these figures we use dimensional quantities, allowing
a simpler comparison to experiments.  As discussed above, we observe
strong decrease of the instability growth rate with respect to the
magnetic field, Fig.~\ref{fig:Smax}.  The two peaks correspond to
quantum tunneling at resonant magnetic fields.  As follows from
Fig.~\ref{fig:Smax}, the instability is the strongest and can be
observed in the region of very small fields.  On the other hand, in
Fig.~\ref{fig:cutoff} we see that the wavelength
$\lambda_{\mathrm{max}}$ corresponding to the maximal growth rate can
be rather high in that range of magnetic field.

It is important to emphasize that the scaling parameters $\kappa_0$
and $\Gamma_0$ are difficult to measure directly.  It is also
difficult to estimate these quantities using theoretical models.  For
instance, values of $\kappa_0 = \num{3e-2} T^{-13/3}$,
$\num{1.5e-5}$, and $\num{3e-9} T^{13/3}$, in units of
\si[per-mode=symbol]{\meter\squared\per\second}, were found using
different heat transfer models~\cite{Velez_PRB_2014}. Here, we used
$\Gamma_0 = \SI{1e7}{\per\second}$ and $\kappa_0 =
\SI[per-mode=symbol]{0.04}{\meter\squared\per\second}$ in
Figs.~\ref{fig:Smax} and \ref{fig:cutoff}.  Our analysis could be used
to measure the product $\kappa_0\Gamma_0$, as the instability of the
magnetic deflagration can be detected in two ways: (i) direct
observation, using magneto-optical
imagining~\cite{Villuendas_EPL_2008}, where the planar front becomes
corrugated with a parabola-like shape as predicted by numerical
simulations~\cite{Jukimenko_PRL_2014,Garanin_PRB_2013}; (ii) the
instability can be detected by measuring deflagration speed, as a
curved front propagates faster then a planar
one~\cite{Jukimenko_PRL_2014}.

A stationary propagation of the curved front was predicted and
explained within non-linear theory in
Ref.~\cite{Jukimenko_PRL_2014}. Similar results were obtained using
direct numerical simulation in Ref.~\cite{Jukimenko_PRL_2014} and
Ref.~\cite{Garanin_PRB_2013}. Importantly, such an instability occurs
regardless of the presence of resonances. Meanwhile, turbulent
propagation of the front observed in Ref.~\cite{Garanin_PRB_2013}
occurs when the field is resonant (where the theoretical model
predicts nearly infinitely fast relaxation). Propagation of the front
at resonant field strengths (when the relaxation rate is extremely
high) must be taken as a separate problem and is not considered here.
Perturbations with different $k$ have different growth rates, see
Eq.~(\ref{eq:disp1}). Therefore, the perturbation with highest growth
rate $\sigma$ develops faster and leads to a stationary, curved front
(see the numerical simulations in Ref.~\cite{Jukimenko_PRL_2014}).

The growth rate stands for the characteristic time needed for the
instability to develop from a planar front to a stationary, curved
front.  For a possible observation of the instability, this time
should be much smaller than the time of propagation of the magnetic
deflagration front within the crystal.  In particular, recent
experiments~\cite{Velez_PRB_2014} were performed with a sample size of
$1.6\times \SI{0.4}{\milli\meter}$ with $B_z=\SI{0.4}{\tesla}$,
corresponding to a typical time of about \SI{6.4e-3}{\second}.
Fig.~\ref{fig:Smax} predicts $1/\sigma = \SI{2e-4}{\second}$ for such
a field magnitude, hence the instability might have enough time to
develop.  However, the wavelength where the growth rate is maximal is
larger than the sample width, depicted as a dashed line in
Fig.~\ref{fig:cutoff}, which leads to a decrease of the instability
growth rate, since the wavelength of the perturbation cannot exceed
the width of the sample.  In addition, this means that the front may
be only slightly curved.

Under restrictions of the size of the sample, the resonant magnetic
field makes the observation of the magnetic instability more
plausible.  We see that the resonance enhances the instability growth
rate in two ways.  First it increases the growth rate
$\sigma_{\mathrm{max}}$, as demonstrated by the peaks in
Fig.~\ref{fig:Smax}.  Second, it decreases the cutoff wavelength and
the wavelength at the maximum (Fig.~\ref{fig:cutoff}), such that the
estimated values are within the range of the dimensions of the
crystal.  In addition to that, the amplitude of the resonance can be
increased significantly by applying a transverse magnetic field,
effectively increasing $\Gamma_0$.  If the instability becomes strong
enough, acceleration of the deflagration front can create a weak shock
wave ahead of the front, which might lead to the
deflagration-to-detonation transition~\cite{Modestov_PRL_2011}.  In
the weak detonation regime, the front propagates at the speed of
sound, and such a spin reversal phenomena has already been observed
experimentally near a resonance~\cite{Decelle_PRL_2009}.

\section{Conclusion}
\label{sec:conclusion}

We have studied the front instability in magnetic deflagration and
found that it behaves in a similar fashion to the Darrieus-Landau
instability in combustion.  That is, in the limit of an infinitely
thin front, the instability has a positive growth rate at all
wavelengths. The dispersion relation of the growth rate as a function
of the instability wave number is also similar to that of the
Darrieus-Landau instability.  The case of a finite-width front was
explored numerically, and we found that the instability should be
observable in current experimental setup, in particular close to a
tunneling resonance, the latter resulting in a smaller value of the
wavelength at which the instability will have the highest growth rate.

Analysing the effect of the direction of propagation of the front with
respect to the easy axis, we showed that the instability would not
grow for a perpendicular front.  We suggest two different experimental
setups (a) and (b), see Fig.~\ref{fig:linear_geometry}. Theory
predicts an unstable front in case (a), resulting in a faster
propagation of the front as result, and a stable front in case (b). By
comparing velocities of the magnetic deflagration of these two
different geometries one can verify the presence of the instability.

Signatures of the presence of the instability might also explain some
previous experimental results.  For instance, for strong longitudinal
fields, the velocities recorded in the experiments are higher than the
theoretical predictions~\cite{Velez_PRB_2014}, which could be
explained by the effect of the instability on front speed.  (Note that
while the front would be curved due to the instability, the
propagation speed of the curved front will also be
steady.~\cite{Jukimenko_PRL_2014}) Likewise, the front broadening also
observed in Ref.~\cite{Velez_PRB_2014} could be explained by the front
instability.

Finally, we showed that there is a relation between the instability
and  the front velocity $U_f$, diffusion constant $\kappa_0$, and the thermal
relaxation rate $\Gamma_R$.  Experimental studies of the instability
could help measure the values of the latter two parameters.

\begin{acknowledgments}
  Financial support from the Swedish Research Council (VR) and the
  Faculty of Natural Sciences, Ume{\aa} University is gratefully
  acknowledged.
\end{acknowledgments}


%

\end{document}